\documentclass[journal]{IEEEtran}
\usepackage[T1]{fontenc}
\usepackage{newtxtext}
\usepackage{booktabs}
\usepackage{amsmath}
\usepackage{listings}
\usepackage{xcolor}
\usepackage{graphicx}
\usepackage{url}
\usepackage{float}
\usepackage{orcidlink}

\definecolor{mlirkw}{rgb}{0.13,0.29,0.53}   % snn ops / structural keywords
\definecolor{mlirty}{rgb}{0.10,0.50,0.30}   % types
\definecolor{mlircm}{rgb}{0.42,0.46,0.50}   % comments
\definecolor{mlirbg}{gray}{0.975}           % background

\lstdefinelanguage{MLIR}{
  alsoletter={.},
  morekeywords={snn.linear,snn.rescale,snn.cubalif,snn.cubali,snn.lif,snn.li,
                ins,out,state,bias},
  morekeywords=[2]{memref,i8,i16,i32,i64,f32,f64,index},
  sensitive=true,
  morecomment=[l]{//},
}

\lstdefinestyle{mlir}{
  language=MLIR,
  backgroundcolor=\color{mlirbg},
  basicstyle=\ttfamily\footnotesize,
  keywordstyle=\color{mlirkw}\bfseries,
  keywordstyle=[2]\color{mlirty},
  commentstyle=\color{mlircm}\itshape,
  showstringspaces=false,
  columns=fullflexible,
  keepspaces=true,
  breaklines=true,
  breakatwhitespace=true,
  frame=single,
  framesep=4pt,
  rulecolor=\color{gray!45},
  xleftmargin=4pt,
  xrightmargin=4pt,
  literate={→}{{$\rightarrow$}}1 {²}{{\textsuperscript{2}}}1
}

\ifCLASSINFOpdf

\else

\fi

\begin{document}

\title{SNN-MLIR: An MLIR Dialect for Compiling Neuromorphic SNNs from NIR to Bare-Metal C }

\author{
    Alejandro~Garc\'ia Gener\,\orcidlink{0009-0000-4110-4138}\IEEEauthorrefmark{1},
    \'Alvaro~Roll\'on de Pinedo\,\orcidlink{0000-0002-9823-5689}\IEEEauthorrefmark{1},
\thanks{\IEEEauthorrefmark{1}INTERA-Group, Barcelona, Spain
\\ (e-mail: alejandro.garcia@intera-group.com,  alvaro.rollon@intera-group.com)}
}

\maketitle

\begin{abstract}
Spiking neural networks (SNNs) are increasingly trained in a wide range of frameworks (\textit{snnTorch}, \textit{Lava}, \textit{Norse}, and others) each with its own model format. The Neuromorphic Intermediate Representation (NIR) addresses this fragmentation by providing a common, framework-independent format for exchanging trained SNN models. NIR solves the exchange problem, but it stops there. It provides a description of a network, not a path to running one. Each backend is still left to implement deployment on its own, with no shared, transformable compiler representation in between. This paper presents \textit{snn-mlir}, an out-of-tree MLIR dialect for SNNs together with a NIR→MLIR→C compilation bridge. The dialect provides a small set of type-polymorphic operations that work identically on floating-point (f32/f64) and quantized (i8/i32) data, so a single intermediate representation serves both simulation and hardware-oriented deployment. A Python front end reads any NIR file and emits dialect IR, automatically inserting rescaling operations to keep quantization scales consistent across layers. A reference lowering pass converts the dialect to standard \textit{linalg} and \textit{arith} operations, from which the toolchain produces self-contained, dependency-free C11 code that compiles and runs on any C-capable CPU or embedded target. We evaluate numerical fidelity against reference outputs, portability across CPU targets, and the cost of quantization. The current scope is feedforward, fully-connected networks with a CPU backend. \textit{snn-mlir} is released as open source under the \textit{Apache-2.0} license with LLVM-exception and it is already available on Github.
\end{abstract}

\begin{IEEEkeywords}
Spiking neural networks, neuromorphic computing, MLIR, NIR, compilers, quantization, edge deployment.
\end{IEEEkeywords}

\IEEEpeerreviewmaketitle

\section{Introduction}

\IEEEPARstart{S}{piking} neural networks have moved from a neuroscience modeling tool to a practical option for low-power inference on edge and embedded hardware. Their event-driven, stateful computation maps naturally onto neuromorphic processors and, increasingly, onto conventional microcontrollers operating under tight energy and memory budgets. This growing interest has produced a healthy ecosystem of training frameworks (snnTorch~\cite{eshraghian2021training}, Lava~\cite{lava2021}, Norse~\cite{norse2021}, Nengo~\cite{dewolf2020nengo} Rockpool~\cite{rockpool}, Sinabs~\cite{sinabs}) each capable of producing trained models. It has also produced a familiar problem: every framework speaks its own language, and a model trained in one is not readily usable in another, let alone on a target device. 

The Neuromorphic Intermediate Representation (NIR) ~\cite{NIR2024} was introduced to solve exactly this exchange problem. NIR  defines a small, framework-independent vocabulary of neuron and synapse primitives (leaky integrate-and-fire variants, linear and affine connections) and lets any framework export a trained network into a common graph format. NIR has been widely adopted precisely because it draws the boundary cleanly: it describes what a network is, deliberately leaving deployment out of scope. That is the right decision for an exchange format, but it leaves a gap. Once a model is in NIR, the path to a running executable is still undefined, and in practice each backend re-implements that path from scratch hand-writing inference loops, target-specific code, and ad hoc quantization and runtimes. There is no shared representation that can be inspected, transformed, and progressively lowered the way a compiler would treat any other program. 

\begin{figure*}[!t]
  \centering
  \includegraphics[width=\textwidth]{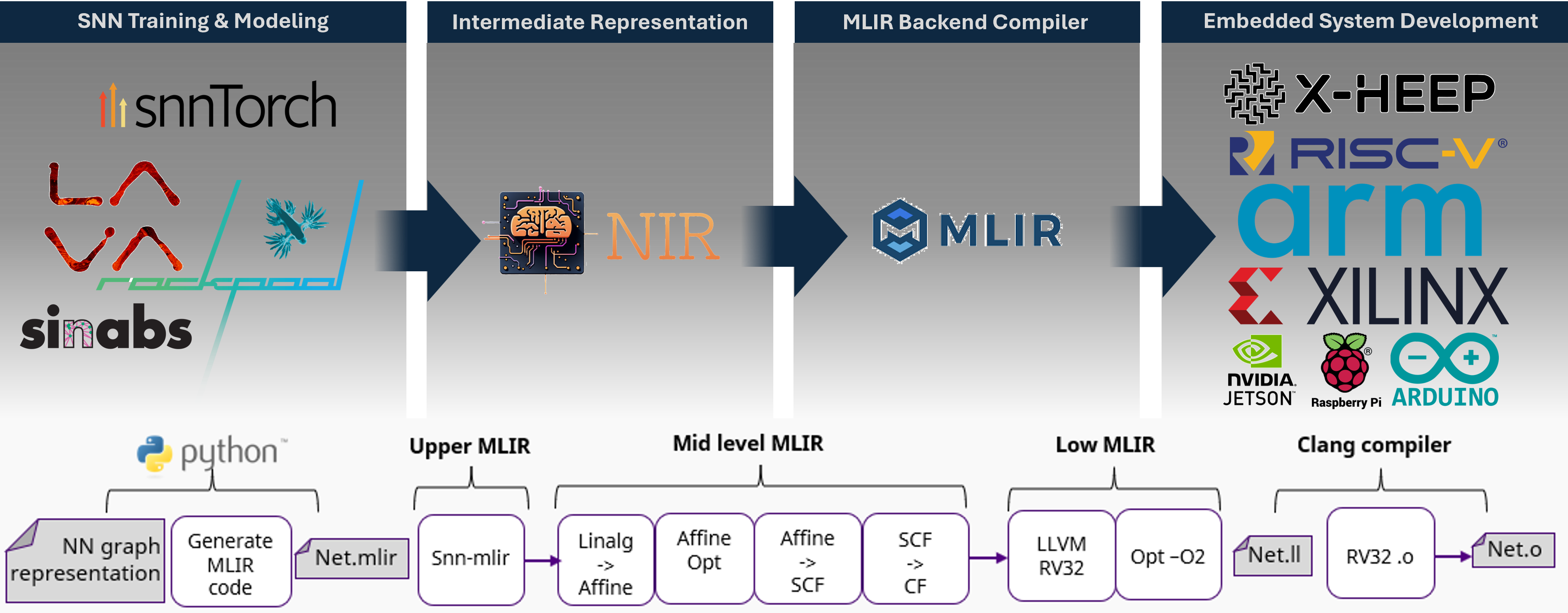}
  \caption{End-to-end compilation flow of snn-mlir. A trained SNN, exported from
  any NIR-compatible framework, passes through the Python front end into the SNN
  dialect, is lowered to standard MLIR and LLVM IR, and compiled to a
  self-contained C11 binary that runs on any C-capable target.}
  \label{fig:flow}
\end{figure*}

Our position is that SNNs need a proper compiler intermediate representation, and that MLIR is the right substrate to build it on. MLIR ~\cite{mlir} is designed for exactly this kind of problem: it supports custom dialects, explicit typing, and progressive lowering from a domain-specific representation down to standard operations and ultimately machine code. By expressing SNNs as a first-class MLIR dialect, we make the network a real compiler object, one that standard tooling can verify, optimize, and re-target, rather than an opaque graph that each backend interprets on its own. 

This paper presents snn-mlir, an out-of-tree MLIR dialect for SNNs and a compilation bridge that takes a NIR model all the way to self-contained C code. Our contributions are:

\begin{enumerate}
    \item \textbf{A type-polymorphic SNN dialect}. A compact set of operations models the CUBA-LIF neuron family and its linear synapses. Each operation works identically on f32 and on quantized i8/i32 data, so simulation and deployment share one IR and one set of ops, with no parallel float and integer op sets to maintain. 

    \item \textbf{A NIR bridge with automatic quantization-scale alignment}. A Python front end reads any NIR graph and emits dialect IR. In quantized mode it inserts rescaling operations automatically to align the differing fixed-point scales between linear layers and neuron dynamics, a mechanism that has no NIR counterpart and that removes a common source of silent error in hand-written quantized inference.

    \item \textbf{A reference CPU lowering}. A conversion pass rewrites dialect operations into standard \textit{linalg} and \textit{arith} operations, which the existing MLIR and LLVM toolchain compiles to LLVM IR and then to machine code. The emitted C runtime is standard, dependency-free C11, compilable by any C-capable target.

    \item \textbf{An evaluation of fidelity and portability}. We verify numerical correctness against reference outputs, demonstrate that the same IR re-targets across multiple CPU and embedded targets, and quantify the accuracy-versus-footprint trade-off of quantization. 
\end{enumerate}

This work deliberately targets a well-defined subset. snn-mlir currently supports feedforward, fully-connected network topologies (linear-chain graphs without branching or recurrence) operating on one-dimensional activation vectors with a batch size of one, and provides a single reference CPU backend. These choices keep the system small and verifiable, but they are also real limitations: Convolutional synapses, branching and recurrent graphs or batched inference are not yet supported. We discuss these constraints in detail later and outline how the design is positioned to lift them. We frame the result not as a faster alternative to GPU inference, but as a portable, dependency-free deployment path that any compatible SNN, from any framework, can reach through NIR.

\section{Background \& related work}
\subsection{SNNs and the Cuba-LIF family}

Unlike the static, real-valued activations of conventional artificial neural networks, spiking neural networks compute over time. A neuron maintains internal state that integrates incoming signals across discrete timesteps, and it communicates with downstream neurons through spikes (binary, all-or-nothing events) rather than continuous values. This temporal, stateful, and largely binary style of computation is what makes SNNs attractive for low-power hardware: work happens only when spikes occur, and the data moving between layers are sparse. 

The dominant neuron model in practice is the leaky integrate-and-fire (LIF)~\cite{gerstner2014neuronal} family. A LIF neuron accumulates input into a membrane potential that decays ("leaks") toward rest each timestep. When the potential crosses a threshold, the neuron emits a spike and resets. The current-based variant (CUBA-LIF) refines this with a second state variable: incoming spikes first charge a synaptic current, which then drives the membrane voltage, giving the neuron two coupled leaky states instead of one. Removing the leak (setting the decay to one) yields the non-leaky integrate-and-fire case (IF), and removing the threshold yields a pure integrator (I) whose continuous membrane potential is read out directly (useful as a final regression or readout layer). These four behaviors (single versus two-state, and spiking versus integrating) span the operations our dialect needs to model, and they are exactly the primitives the NIR standard exposes. 

For a compiler audience, the salient point is that these neurons are not point-wise functions. Each carries state that must persist across timesteps, a threshold, reset and control decision, and a sharp distinction between binary spike outputs and continuous voltage outputs. A representation that flattens them into anonymous arithmetic loops discards precisely the structure a compiler would want to reason about.

\subsection{NIR as an exchange format}
The Neuromorphic Intermediate Representation addresses the fragmentation of the SNN ecosystem by defining a framework-neutral graph format with a fixed vocabulary of node types (linear and affine synapses, and the LIF/CUBA neuron family among them). A network trained in one framework can be exported to NIR and read back in another, and a growing list of simulators (including LAVA, SNNTorch, Norse, Nengo, Rockpool, and Sinabs \cite{eshraghian2021training,lava2021,norse2021,dewolf2020nengo,rockpool,sinabs}) support it as an export target. By consuming NIR, a downstream tool inherits all of these frameworks at once, rather than maintaining an individual importer per framework. 

NIR is also deliberately broad in physical scope. It describes both continuous-time analog and discrete-time digital neuron models, parameterizing them with continuous quantities such as membrane and synaptic time constants, leak, and threshold. This generality is a strength for exchange, but it means a NIR graph is a mathematical description of a network, not an executable artifact. NIR fixes the semantics of a model. It intentionally says nothing about how that model is discretized, quantized, scheduled, or compiled for a particular target. Turning a NIR graph into running code is left entirely to the consumer and that is the gap this work fills. 

\subsection{MLIR and Progressive Lowering}
MLIR is a compiler infrastructure organized around dialects: Self-contained sets of operations, types, and attributes that model a particular domain at a particular level of abstraction. A program is expressed in high-level, domain-specific dialects and then progressively lowered (through a sequence of rewrite passes) toward successively lower-level dialects, until it reaches a form (such as the LLVM dialect) from which machine code can be generated. Every operation, at every level, is first-class IR: it can be verified, analyzed, and transformed by standard passes, and dialects can be mixed within a single module. This makes MLIR a natural home for a new domain abstraction. A custom dialect can capture domain semantics explicitly at the top of the pipeline, then reuse the entire existing ecosystem of lower-level dialects and code generators beneath it, rather than building a compiler stack from scratch. 

\subsection{Related Work}
Most existing routes from a trained SNN to a deployed one are framework-specific. A given simulator typically ships its own deployment or code-generation path, tied to its own model format and often to a particular hardware target. These paths work well within their ecosystem but do not compose: a model must commit to one framework's toolchain, and porting to a different target generally means re-exporting, re-quantizing, and re-implementing the inference loop. NIR removes the format-level barrier between frameworks, but it stops at the model boundary, so each backend still re-implements deployment independently—frequently as hand-written inference loops with target-specific quantization handling.

 A second line of work targets neuromorphic hardware directly, mapping networks onto specific chips through vendor toolchains. Such compilers are valuable but are, by construction, specialized to their target device and programming model. What is missing between these two worlds is a shared, framework-agnostic compiler IR: a representation that is neutral to the source framework, explicit about SNN semantics, and transformable and retargetable like any other compiler program. snn-mlir occupies that layer. It takes the framework-neutral model that NIR provides and gives it a compiler-neutral intermediate representation in MLIR, one IR that can be inspected, optimized with standard tooling, and lowered to a CPU today or to custom hardware tomorrow through the same backend-extension mechanism.

 \section{The SNN dialect}
 The dialect is built on three deliberate choices.

 \begin{itemize}
     \item \textbf{It is memref-based}: Every operand (inputs, weights, neuron state, and outputs) is a memory reference rather than an SSA value. This matches the reality of the domain, where neuron state must persist and be updated in place across timesteps, and it maps cleanly onto the buffers a C runtime or an embedded target actually allocates.  

     \item \textbf{The dialect is explicitly typed}: Every operation carries the element type and shape of its operands in the IR text, so the same operation reads unambiguously whether it is operating on floating-point or integer data.

     \item \textbf{The dialect is designed to be one IR for both simulation and deployment}: The same NIR graph serves as a faithful floating-point reference and as the source for a quantized, hardware-oriented build; there is no separate "deployment IR" that can drift out of sync with the reference.
 \end{itemize}

The guiding idea behind all three is that a spiking network should be a first-class compiler object. A neuron layer is one operation that carries its own decay constants and threshold, not an opaque nest of arithmetic. Standard MLIR passes can verify it, inspect it, and rewrite it, and the network stays analyzable and retargetable all the way down.

The dialect defines six operations, covering linear synapses (with and without bias), rescaling, and the four neuron behaviors of the CUBA-LIF family (TABLE~\ref{tab:ops}). Each operation has a human-readable assembly format, so the IR is directly inspectable and hand-editable. As an example, the CUBA-LIF neuron layer makes its full structure explicit: The input current, the two in-place state buffers, the spike output, and the dynamics parameters as attributes.

\begin{table*}[!t]
\centering
\caption{SNN Dialect Ops}
\label{tab:ops}
\resizebox{\textwidth}{!}{%
\begin{tabular}{llll}
\toprule
Op & States & Output & Summary \\
\midrule
\texttt{snn.linear}  & ---              & \texttt{f32}/\texttt{i32} & Matrix-vector synapse layer + optional bias \\
\texttt{snn.rescale} & ---              & \texttt{i32}              & Per-edge requantization shift to align quantization scales \\
\texttt{snn.cubalif} & Current, voltage & \texttt{f32}/\texttt{i8}  & Current-based leaky integrate-and-fire: two states, threshold and reset \\
\texttt{snn.cubali}  & Current, voltage & \texttt{f32}/\texttt{i32} & Current-based leaky integrator: two states, continuous voltage output \\
\texttt{snn.lif}     & Voltage          & \texttt{f32}/\texttt{i8}  & Leaky integrate-and-fire: single state, threshold and reset \\
\texttt{snn.li}      & Voltage          & \texttt{f32}/\texttt{i32} & Leaky integrator: single state, continuous voltage output \\
\bottomrule
\end{tabular}}
\end{table*}

The operation reads as exactly what it computes: a 256-neuron CUBA-LIF layer that decays its synaptic current and membrane voltage each timestep, fires when the voltage crosses the threshold, and resets. The state operands are read and written in place, capturing the temporal nature of the neuron directly in the IR.

\begin{figure}[H]
\begin{lstlisting}[style=mlir]
snn.cubalif ins(%input) state(%current, %voltage) out(%output)
    { cur_decay_float = 0.95 : f64,
      vol_decay_float = 0.90 : f64,
      threshold_float = 1.0  : f64 }
    : memref<256xf32>, memref<256xf32>, memref<256xf32> -> memref<256xf32>
\end{lstlisting}
\caption{A 256-neuron CUBA-LIF layer in floating-point. The \texttt{state(...)} operands are updated in place each timestep; dynamics parameters are carried as \texttt{f64} attributes on the op itself.}
\label{fig:snn-dialect}
\end{figure}

The central design decision of the dialect is that a single operation works on both floating-point and quantized data. \textit{snn.cubalif} operating on \textit{memref<256xf32>} and the same operation operating on \textit{memref<256xi32>} state with an i8 spike output are the same op, not two members of a parallel float and integer op set that must be kept in step.

This is achieved by carrying both parameter representations on the operation and selecting between them by operand type. Each neuron op holds its dynamics parameters twice: as real-valued f64/f32 attributes (\textit{cur\_decay\_float}, \textit{threshold\_float}, etc.) and as Q12 fixed-point i32 attributes (\textit{cur\_decay\_int}, \textit{threshold\_int}, etc. together with the \textit{d\_scale}). The lowering reads only the set that matches the operand element type and the unused set keeps its default. The quantized version of the layer above is the same operation with integer operands and the integer attribute set:

\begin{figure}[H]
\begin{lstlisting}[style=mlir]
snn.cubalif ins(%input) state(%current, %voltage) out(%output)
    { d_scale = 12 : i32, 
      cur_decay_int = 3891 : i32, 
      vol_decay_int = 3686 : i32, 
      threshold_int = 4096 : i32}
    : memref<256xi32>, memref<256xi32>, memref<256xi32> -> memref<256xi8>

\end{lstlisting}
\caption{The same \texttt{snn.cubalif} op in Q12 integer mode. Only the operand types and attribute set change; the op, its verifier, and its lowering pattern are shared with the float case.}
\label{lst:snn-dialect}
\end{figure}

The payoff is concrete. A network can be lowered in floating point to produce a bit-faithful reference and lowered again in quantized form for an embedded target, from one dialect, one op set, and one set of verifiers and passes. There is no second code path to maintain and no risk that the float and integer semantics diverge. 
The neuron ops split along the output type:

\begin{itemize}
    \item \textbf{Spike-output ops} (\textit{snn.cubalif}, \textit{snn.lif}) apply a threshold and emit binary activations and reset on firing.
    \item \textbf{Voltage-output ops} (\textit{snn.cubali}, \textit{snn.li}) have no threshold and emit the continuous membrane potential directly; they serve as the final readout layer in regression or readout networks.
\end{itemize}

Encoding this distinction in the op set, rather than as a flag, lets a backend treat the two cases differently without re-deriving the network's intent. The integrate-and-fire variants are not separate ops at all: Setting the decay to one disables the exponential leak, so \textit{snn.IF} and \textit{snn.CubaIF} are simply the leaky neurons with unit decay.

Finally, the dialect treats its restrictions as explicit, verifier-enforced design choices rather than silent assumptions. Every op requires one-dimensional activation vectors (\textit{memref<Nxf32>} or \textit{memref<Nxi32>}): neuron populations are flat arrays, not spatial feature maps. Each op carries a verifier (\textit{hasVerifier = 1}) that checks operand ranks and type consistency, so feeding a two-dimensional feature map produces a clear diagnostic at verification time instead of a silent miss compilation downstream. This keeps the compiler honest and machine-checkable, and it gives a well-defined surface against which future extensions, such as a convolutional synapse op, can be added.

\section{Quantization \& Discretization}
The quantization scheme is power-of-two (shift-based), inspired by Lava's approach ~\cite{lava2021} . Using powers of two means every rescaling is a bit-shift rather than a multiply-and-divide (cheap and exact on integer-only embedded hardware) while keeping accuracy loss low.

\textbf{Weights: i8}. For each synapse layer, a single per-layer scale is chosen so the largest-magnitude weight maps near the i8 range without overflowing it:

\begin{align}
  \text{ratio}   &= \min\!\left( \left| \frac{127}{\max_w} \right|,\ \left| \frac{128}{\min_w} \right| \right) \\
  w_{\text{scale}} &= \left\lfloor \log_2 (\text{ratio}) \right\rfloor \\
  q_w            &= \operatorname{round}\!\left( w \cdot 2^{\,w_{\text{scale}}} \right)
\end{align}

The exponent w\_scale (not the factor) is recorded as an attribute on \textit{snn.linear}, so the scale is recoverable downstream. A bias, if present, is quantized with the same scale and stored as \textit{int32}.

\textbf{Neuron state: \textit{Q12} fixed point}. Neuron parameters (decays, threshold) are quantized to a fixed Q12 format, i.e. scaled by $2^{12}$, and emitted as integer attributes on the neuron op. The state arrays become i32 and spike outputs become i8 (0/1). The neuron's scale exponent is \textit{d\_scale = 12}

\textbf{The problem, rescaling}. These two scales are chosen independently and almost never match. A \textit{snn.linear} produces values at scale $2^{w\_scale}$, but the following neuron op expects its input at the neuron's scale $2^{d\_scale}$. Feeding one into the other unaltered would silently corrupt the dynamics. The front end resolves this by inserting a synthetic \textit{snn.rescale} operation on every synapse-neuron edge that shifts by exactly the difference of the two exponents:

$$ shift = d\_scale - w\_scale $$

A positive shift is a left shift, a negative one an arithmetic right shift; the input is sign-extended to i32 before shifting. This operation exists only in quantized output and has no NIR equivalent. It is created by the export layer once both neighboring scales are known, replacing what would otherwise be a floating-point rescale with a single integer shift.

A working example. Consider a linear layer whose weights range over roughly $[-0.7, 0.9]$. The scale search gives 

$$ratio = \min(127/0.9, 128/0.7) \approx \min(141, 183) = 141$$ 

$$w\_scale = floor(\log_2(141)) = 7$$

Weights are stored as $round(w \cdot 2^7)$, mapping $0.9$ to $115$ and $-0.7$ to $-90$, safely inside i8. The layer therefore emits its result at scale $2^7$. The CUBA-LIF neuron that follows works in Q12, so $d\_scale = 12$. The front end inserts a rescale with $shift = 12 - 7 = 5$, left-shifting the synapse output to align it with the neuron's fixed-point format before the dynamics run. In dialect IR this is three consecutive operations:

\begin{figure}[H]
\begin{lstlisting}[style=mlir]
// int8 synapse, output at scale 2^7
snn.linear ins(%x, %w) out(%syn) { w_scale = 7 : i64 }
    : memref<200xi8>, memref<256x200xi8> -> memref<256xi32>

// align 2^7 -> 2^12 with a single left shift of 5
snn.rescale ins(%syn) out(%resc) { w_scale = 7 : i64, d_scale = 12 : i64 }
    : memref<256xi32> -> memref<256xi32>

// CUBA-LIF dynamics in Q12, i8 spike output
snn.cubalif ins(%resc) state(%cur, %vol) out(%spk)
    { d_scale = 12, cur_decay_int = 3891, vol_decay_int = 3686, threshold_int = 4096 }
    : memref<256xi32>, memref<256xi32>, memref<256xi32> -> memref<256xi8>
\end{lstlisting}
\caption{Quantized synapse $\rightarrow$ rescale $\rightarrow$ neuron chain in the
SNN dialect: the linear layer emits at scale $2^{7}$, \textit{snn.rescale} shifts
left by $5$ to reach the neuron's $Q12$ scale, and \textit{snn.cubalif} runs the
integer dynamics.}
\label{lst:quant-chain}
\end{figure}

The result is a fully integer pipeline whose inter-layer scales are provably consistent by construction, generated automatically from a float NIR model with a single flag. The mechanism is small, but it removes one of the most common and hardest-to-debug sources of error in hand-written quantized SNN inference, a silent scale mismatch between a synapse and the neuron it drives.

\section{Lowering and code generation}
The dialect is the top of the pipeline. Everything below it reuses standard MLIR and LLVM infrastructure. This section describes the reference lowering from snn ops to standard dialects, the CPU pipeline that carries it to LLVM IR, and the C runtime that wraps the result into a self-contained, dependency-free executable.

The reference lowering, \textit{SNNToLinalg}, is a conversion pass that rewrites each snn operation into standard \textit{linalg} and \textit{arith} operations, the level at which the rest of the MLIR ecosystem can take over. The pass installs one rewrite pattern per op and marks the snn dialect illegal, so a successful run leaves no snn operation behind.

The key point is that the pass dispatches on operand element type, which is what makes the dialect's type polymorphism pay off at the code-generation level. For \textit{snn.linear}, the floating-point path lowers to a \textit{linalg.matvec}; the quantized path lowers to a \textit{linalg.generic} whose body sign-extends the i8 operands and accumulates into the wider i32 output, implementing the integer matrix-vector product with the correct widening. The neuron ops lower similarly, expanding their two-line update (decay, integrate, threshold-and-reset) for spiking neurons; decay and integrate for integrators into \textit{arith} operations over the activation vector, and reading the float or int parameter attributes according to the operand type. Because the float and integer cases share one op and one pattern, the two code paths cannot drift apart. They are the same lowering, specialized by type.

The network itself is emitted as a single MLIR function, carrying the LLVM attribute. This function advances the network by exactly one timestep, taking the weight buffers and the neuron state buffers as \textit{memref} arguments and updating the state in place. Driving it across time is left to the runtime, which keeps the compiled kernel small and stateless from the caller's point of view. Then, the \textit{snn-opt}, the project's standalone opt tool, which registers the dialect and the lowering, applies \textit{SNNToLinalg}.

From here, lowering to an executable is a straight, file-free chain of standard tools. The result is then handed to stock \textit{mlir-opt}, which converts \textit{linalg} to affine loops, runs standard cleanup (loop-invariant code motion, CSE, canonicalization), and lowers progressively through affine, scf, cf and the memref/arith/func dialects down to the LLVM dialect. Finally \textit{mlir-translate} emits textual LLVM IR. From this point, any LLVM-based compiler produces an object file or executable. 

Every stage below snn-opt is an unmodified upstream tool. The project contributes only the dialect and its one lowering pass, and inherits the entire optimization and code-generation stack beneath them.

\section{Evaluation}

We evaluate snn-mlir on four questions: 
\begin{itemize}
    \item \textbf{Fidelity}: Does the compiled output reproduce the
source framework's behavior?. 
    \item \textbf{Portability}: Does the same IR deploy unchanged
across different targets?. 
    \item \textbf{Baseline}: How does the dependency-free C binary
compare to running the network in its framework?.
    \item \textbf{Trade-off}: What does
quantization cost and buy?.
\end{itemize}
The evaluation is validation-oriented:
the goal is a correct, deployable path, not one faster than a GPU.

\subsection{Validation Setup}

We use the two example networks shipped with the project, each exported to NIR
from a different framework:

\begin{itemize}
  \item \textbf{snn\_oxford}: A two-layer current-based LIF network trained
        with LAVA-DL from the examples documentation~\cite{lava_oxford}, evaluated over the first 100 timesteps:  
        
        Linear(200,256) $\rightarrow$
        CubaLIF(256) $\rightarrow$ Linear(256,200) $\rightarrow$
        CubaLIF(200)
  \item \textbf{snntorch}: A four-layer network exported from snnTorch, evaluated over 25
        timesteps :
        
        Linear(784,256) $\rightarrow$ LIF(256) $\rightarrow$
        Affine(256,10) $\rightarrow$ CubaLIF(10)
\end{itemize}

Each network is compiled through the full pipeline in both floating point and quantized modes and compared against a reference spike trace from the original framework. 
The fidelity metric is per-cell spike
equality, the fraction of output cells (timesteps $\times$ neurons) matching
the reference together with the total spike count. Timing is reported in CPU
cycles per timestep, which is frequency-independent and therefore comparable
across targets.

Two targets are used. The x86 host is a 13th Gen Intel Core i7-13620H
(16 logical cores, x86\_64) running Ubuntu 24.04 under WSL2 at an effective
2.04~GHz, with LLVM/MLIR 22.x and clang (this is a full operating-system
environment). The embedded target is an X-HEEP 32-bit RISC-V
core on FPGA, running bare-metal at a 15~MHz
prototyping clock, with the C runtime cross-compiled. 
Only the quantized i8 build is deployed to X-HEEP,
as is appropriate for an integer-only embedded core.

\subsection{Numerical Fidelity}

In \textbf{float mode the compiled output is bit-exact with the reference}
for both networks: every output cell matches and the total spike counts are
identical (Table~\ref{tab:fidelity}). This confirms that the dialect, its
lowering, and the C runtime reproduce the source framework's dynamics exactly,
with no drift across the discretized timestep loop.

In \textbf{quantized int8 mode} the output is no longer bit-exact, as expected
from the precision reduction, but the most important result is consistency
across targets: the x86 and the X-HEEP RISC-V runs produce 99.585\% accuracy on 
snn\_oxford and 92.000\% on snntorch, the same
mismatched cells and the same spike counts on both machines. Because the
quantization is integer-only and shift-based, the compiled network is
deterministic across two entirely different instruction sets. The embedded
target is not an approximation of the desktop result but exactly equal to it.
The residual error versus the float reference is network-dependent: negligible
for snn\_oxford ($-0.4\%$, a net difference of one spike over 100 timesteps) and
modest for snntorch ($-8\%$).

\begin{table}[H]
\centering
\caption{Numerical fidelity, across precisions and targets.}
\label{tab:fidelity}
\resizebox{\columnwidth}{!}{%
\begin{tabular}{llccc}
\toprule
Network & Target / mode & Spikes (act./ref.) & Mismatches & Bit acc. \\
\midrule
snn\_oxford & x86, float32 & 329 / 329 & 0 / 20{,}000  & 100.000\% \\
snn\_oxford & x86, int8    & 328 / 329 & 83 / 20{,}000 & 99.585\%  \\
snn\_oxford & X-HEEP, int8 & 328 / 329 & 83 / 20{,}000 & 99.585\%  \\
\midrule
snntorch    & x86, float32 & 178 / 178 & 0 / 250  & 100.000\% \\
snntorch    & x86, int8    & 180 / 178 & 20 / 250 & 92.000\%  \\
snntorch    & X-HEEP, int8 & 180 / 178 & 20 / 250 & 92.000\%  \\
\bottomrule
\end{tabular}}
\end{table}

\subsection{Portability and Footprint}

The same NIR graph that produced the x86 binary cross-compiles,
unchanged, to the X-HEEP RISC-V core, only the C compiler differs. On-target memory usage
(Table~\ref{tab:footprint}) separates cleanly into two parts. The code
and runtime footprint is fixed at $\approx$12.6~KB (9.6\% of the 128~KB
instruction region) and is identical for both networks. It is the
timestep kernel, the \textit{memref} glue, and the stack, none of which depend on the
model. The data footprint scales with the model and is dominated by the
i8 weight matrices: 114~KB for snn\_oxford (43.6\% of the 256~KB data region)
and 172~KB for snntorch (67\%).

That contrast is the practical portability story. Both models fit comfortably 
on a mid-range microcontroller, but bigger SNNs would not fit on small
devices such as an Arduino-class part or a low-end STM32, which offer only tens
of kilobytes of RAM. The limiting factor for deploying a given network is
therefore the model size, set almost entirely by its weights, and this
is exactly what quantization addresses, cutting the weight footprint 
and widening the range of devices a network can reach. The fixed
$\sim$12.6~KB kernel, by contrast, is never the obstacle.

\begin{table}[H]
\centering
\caption{On-target footprint and per-step cost (quantized int8).}
\label{tab:footprint}
\resizebox{\columnwidth}{!}{%
\begin{tabular}{llccr}
\toprule
Network & Target & Code+runtime & Weight+state  \\
\midrule
snn\_oxford & X-HEEP & 12{,}584\,B (9.6\%) & 114{,}416\,B (43.6\%)  \\
snntorch    & X-HEEP & 12{,}584\,B (9.6\%) & 172{,}912\,B (67.1\%)  \\
\bottomrule
\end{tabular}}
\end{table}

\subsection{Baseline Comparison}

To put the compiled binary in context, we compare its
per-inference-step cost against running the same network in its
original Python framework on the same x86 host (Table~\ref{tab:baseline}). The
framework baseline is large, \textbf{173.3M cycles per step} for snn\_oxford and
\textbf{194.0M for snntorch} because each step pays the full cost of the Python
interpreter and the deep-learning runtime: tensor allocation, dispatch, and
library call overhead, repeated on every timestep. The snn-mlir i8 binary runs
the same step in \textbf{651k} and \textbf{1.23M} cycles respectively, two-to-three
orders of magnitude lower ($\approx$266$\times$ and $\approx$157$\times$);
the bit-exact float binary is 213$\times$ and 118$\times$ lower.

This is not a peak-performance claim. For a small network on a CPU, the
framework's per-step time is dominated by fixed interpreter overhead that has
nothing to do with the network's arithmetic, roughly 85--95~ms per step at
2.04~GHz, versus well under a millisecond for the compiled binary. The point is
what that overhead means for embedded deployment. A framework must carry an
interpreter and its dependencies to run a single timestep, whereas the snn-mlir
artifact is a standalone C binary with none of it. Removing that stack is
precisely what makes the network deployable on a bare-metal embedded
target, the same artifact that posts these x86 numbers is the one that runs on
X-HEEP.

\begin{table}[H]
\centering
\caption{Baseline comparison: cycles \emph{per inference step} (x86).
The framework baseline includes \textit{Python} interpreter and runtime overhead.}
\label{tab:baseline}
\resizebox{\columnwidth}{!}{%
\begin{tabular}{lrrrr}
\toprule
Network & \shortstack{snn-mlir\\float32} & \shortstack{snn-mlir\\int8}
        & \shortstack{Framework\\baseline} & \shortstack{Speedup\\(int8)} \\
\midrule
snn\_oxford & 812{,}110     & 651{,}270     & 173{,}303{,}778 & 266$\times$ \\
snntorch    & 1{,}639{,}168 & 1{,}233{,}384 & 194{,}004{,}915 & 157$\times$ \\
\bottomrule
\end{tabular}}
\end{table}

\subsection{Quantization Trade-off}

Table~\ref{tab:quant} isolates what quantization costs and buys on the x86 host,
pairing the accuracy from the fidelity results with speed and size. i8
inference is \textbf{1.25$\times$ faster on snn\_oxford} (0.398 $\rightarrow$
0.319~ms/step) and \textbf{1.34$\times$ faster on snntorch} (0.819 $\rightarrow$
0.612~ms/step), while shrinking the weight data by roughly 4$\times$. The
accuracy cost is small and network-dependent, negligible for snn\_oxford,
modest for snntorch. The power-of-two, shift-based scheme keeps these gains
essentially free of arithmetic overhead. Every resize is a single shift,
not a multiply-and-divide, which is also why the integer build is deterministic
enough to be bit-identical on x86 and RISC-V.

% requires \usepackage{booktabs}
\begin{table}[H]
\centering
\caption{Quantization trade-off (float32 $\rightarrow$ int8, x86).}
\label{tab:quant}
\setlength{\tabcolsep}{4pt}
\footnotesize
\resizebox{\columnwidth}{!}{%
\begin{tabular}{lcc}
\toprule
Metric & snn\_oxford & snntorch \\
\midrule
Accuracy     & 100\% $\rightarrow$ 99.59\%   & 100\% $\rightarrow$ 92.00\%   \\
Time/step    & 0.398 $\rightarrow$ 0.319\,ms & 0.819 $\rightarrow$ 0.612\,ms \\
Speedup      & 1.25$\times$                  & 1.34$\times$                  \\
Weight data  & 1.86\,MB $\rightarrow$ 443\,KB & 3.66\,MB $\rightarrow$ 939\,KB \\
Compression  & 4.20$\times$                  & 3.90$\times$                  \\
\bottomrule
\end{tabular}}
\end{table}

\begin{figure}[H]
  \centering
  \includegraphics[width=\columnwidth]{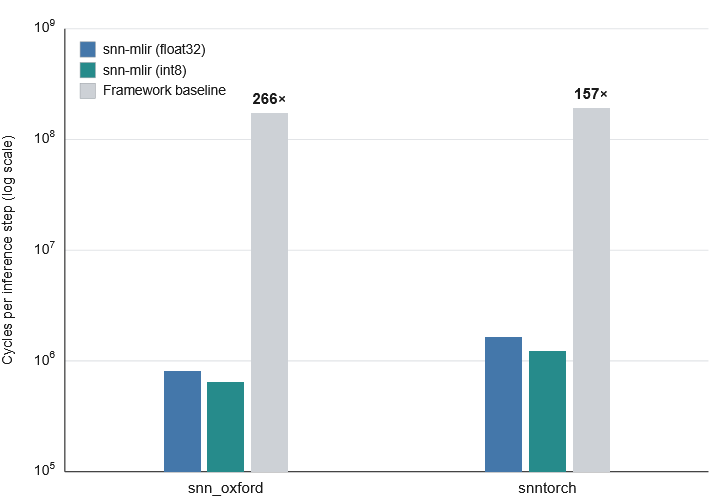}
  \caption{Per-step inference cost on x86 (log scale). The snn-mlir float32
  and int8 binaries run 213--266$\times$ faster than the framework baseline,
  which pays the full Python interpreter and runtime overhead on every timestep.}
  \label{fig:baseline}
\end{figure}

\section{Limitations}

The current implementation covers feedforward, fully-connected SNN topologies. The following are known constraints:

\textbf{No convolutional ops.} NIR nodes such as \texttt{nir.Conv2d}, \texttt{nir.AvgPool2d}, and \texttt{nir.SumPool2d} operate on \texttt{[channels, height, width]} feature maps and have no equivalent SNN op yet. Supporting them requires new ops (e.g., \texttt{snn.conv2d}). The neuron dynamics ops are already rank-agnostic at the lowering level; extending them to $N$-D is straightforward once a convolutional synapse op exists.

\textbf{Linear-chain graphs only.} The Python graph walker follows a single path from input to output. Branching, residual connections, and recurrent edges are not supported.

\textbf{Batch size 1.} There is no batched-inference mode. Each call to the compiled function processes one input sample as it is intended by NIR standard. The objective are embedded stream applications rather than batch training acceleration. 

\textbf{Uniform neuron parameters per layer.} All neurons in a layer share the same decay constants and threshold. Per-neuron parameter arrays are not yet supported.

These constraints are enforced rather than silently assumed. Each one is surfaced at dialect verification time, so they manifest as clear diagnostics rather than incorrect output.

\section{Future Work}

Convolutional synapse ops are the
most immediate extensions. They would lift the 1-D activation restriction and
enable convolutional SNN topologies common in event-driven vision and
time-series workloads. The neuron dynamics ops already lower through
type-dispatch and are not inherently limited to 1-D data; the blocker is the
synapse op, not the neuron dynamics.

Residual and recurrent graph support requires extending the Python graph walker
and IR emission layer to handle fan-out and fan-in edges. The MLIR module format
already accommodates this naturally (a multi-path network is a function with
additional \textit{memref} arguments) so the change is concentrated in the
front end. Recurrent graph support, in particular, would extend snn-mlir to
Liquid State Machines, an important class of reservoir computing SNNs not
covered by the current linear-chain walker.

Neuromorphic hardware backends are a natural next step once a stable dialect
exists. The lowering-pass template established by \textit{SNNToLinalg}
generalizes directly: a new conversion pass targeting a vendor dialect (e.g.,
Intel Loihi~2 via Lava, or SpiNNaker) follows the same one-pattern-per-op
structure, inheriting type-polymorphic dispatch and existing verifiers for free.
We actively welcome open-source contributions of new backends and hardware
architecture targets from the community.

The most strategically important next step, however, is coordination with the
NIR development team to establish a standardized quantization and discretization
methodology across the ecosystem. The current scheme (power-of-two weight
quantization, Q12 neuron state, automatic rescaling) works well in practice,
but it is one approach among several plausible ones. A shared, NIR-level
convention for quantization parameters agreed upon across frameworks and tools
would remove the last remaining per-backend divergence and make quantized SNN
deployment truly interoperable. snn-mlir is committed to contributing to and
adopting whatever standard the community converges on.

\section{Conclusion}

snn-mlir gives spiking neural networks a proper compiler intermediate representation for the first time. A trained SNN, exported from any NIR-compatible framework, enters a type-polymorphic MLIR dialect whose operations carry explicit neuron semantics, are verified by the standard MLIR infrastructure, and lower through a single, auditable pass to dependency-free C11 code that runs unchanged on any C-capable CPU or embedded target. The same IR serves both floating-point simulation and integer deployment, with quantization scales aligned automatically by the front end, removing one of the most common sources of silent error in hand-written quantized SNN inference. Numerical fidelity is bit-exact in floating point and deterministic across architectures in i8. The project is open source under the Apache-2.0 license with LLVM-exception and is available at \url{https://github.com/INTERA-GROUP/snn-mlir}. 

snn-mlir is a starting point, not a finished product. We hope it serves as a
useful foundation and we welcome collaborators, new backends, hardware targets,
and ecosystem integrations. The project is designed to grow with the
neuromorphic community, and every contribution (from a new lowering pass to a
standards discussion) is valued.

\section*{Acknowledgment}

The authors thank the Open Neuromorphics community and the NIR development team for maintaining an open,
framework-neutral SNN exchange format that made this work possible, and the
MLIR and LLVM communities for the infrastructure on which snn-mlir is built.
We also extend our thanks to all future contributors who will help shape,
extend, and improve this project, this work belongs to the community as much
as to its authors. 

\ifCLASSOPTIONcaptionsoff
  \newpage
\fi

\bibliographystyle{IEEEtran}
\bibliography{references}
\end{document}